\documentstyle[preprint,aps]{revtex}
\input psfig
\tighten
\begin{document}
\draft
\title{Robust evolution system for Numerical Relativity}
\author{A. Arbona, C. Bona, J. Mass\'{o} and J. Stela}
\address{Departament de F\'\i sica. Universitat de les Illes
Balears.\\
 E-07071 Palma de Mallorca, SPAIN}
\maketitle
\begin{abstract}
The paper combines theoretical and applied ideas which have been
previously considered separately into a single set of evolution
equations for Numerical Relativity. New numerical ingredients are
presented which avoid gauge pathologies and allow one to perform
robust 3D calculations. The potential of the resulting numerical
code is demonstrated by using the Schwarzschild black hole as a
test-bed. Its evolution can be followed up to times greater than
one hundred black hole masses.
\end{abstract}

\narrowtext
\section{Introduction}
In spite of the valuable work made by many Numerical Relativity
groups all around the world, three-dimensional (3D) gravitational
systems are still challenging (see Ref \cite{Cactus} for a recent
review of the field) . This is not only because the present
generation of computers does not allow yet to perform high
resolution 3D calculations. This is also, or at least we believe
it is, because there are new numerical problems associated with 3D
gravitational calculations which were not recognized in the
previous work with spherically symmetric (1D) or even axially
symmetric (2D) systems. To identify these new problems and to deal
with them requires new developments, both from the theoretical and
from the applied point of view.

The theoretical developments, as presented in Section II, are made
in the framework of the well known 3+1
formalism\cite{Lich,Choquet,ADM}. We propose first to change the
standard set of evolution equations (obtained from the space
components of the four-dimensional Ricci tensor) to the one
obtained from the space components of the Einstein tensor (which
we called 'Einstein system'\cite{PRD97} for obvious reasons).

We will now incorporate to this second-order system the idea of
transforming the momentum constraint into an evolution equation
for a set of three supplementary quantities, which we used in
former works with first-order evolution systems\cite{PRL92,PRL95}.
Although the original idea was devised to obtain hyperbolic
evolution systems, variants of the same approach have been
considered recently\cite{SN,BS,Potsdam} as a tool to improve the
results in specific numerical simulations.

These modifications do change completely the causal structure of
the original system, as discussed in Refs\cite{PRL92,PRL95}. The
ones presented in Section III, on the contrary, are minor changes
from the theoretical point of view but they seem to be crucial
from the computational one. They address the problem of the so
called 'gauge pathologies'\cite{Miguel,MiguelJoan}, which consists
in the explosive behaviour of the quantities associated with the
gauge modes. This problem seldom happened in 1D calculations (at
least for the standard settings) but, in our experience, it turns
out to be the rule in 3D calculations.

A possible remedy to this problem could be trying to fix the
gauge-related degrees of freedom by using the well known maximal
slicing\cite{York72}. This requires, however, to solve a 3D
elliptic equation at every time step, which implies a huge
computational cost, specially in high resolution calculations. But
this is not the only alternative available: the gauge-related
degrees of freedom can be isolated from the rest of the system by
using the well known conformal decomposition of the space
metric\cite{Lich,York71} as in the recent work of Shibata and
Nakamura\cite{SN}, which use the harmonic
slicing\cite{Choquet83,PRD88}, which is much less expensive from
the computational point of view.

Both alternatives, however, are not sufficient by themselves to
allow for long term 3D calculations in strong field situations,
like black hole space-times. Even if one tries to combine both
approaches, the errors in gauge modes show a tendency to grow
without bound which should be tackled using improved
tools\cite{Potsdam}. We have devised numerical counter-terms which
have shown to be very effective in controlling these errors
without modifying the analytical system of equations. As they are
not restricted to maximal slicing, one is then allowed to use
other slicings with a great increase in the code speed. The use of
these counter-terms opens then the door to performing robust 3D
numerical calculations at a level similar to that of previous 2D
or even 1D results.

We present in Section IV our results for the single black hole
case as a test-bed for our finite difference code. We obtain long
term evolution simulations, up to more than one hundred times the
black hole mass.  A convergence analysis shows that the collapse
speed one obtains (the rate at which the numerical grid falls into
the hole) depends on the grid resolution, but it converges to the
well established results for spherically symmetric (1D) black
holes. The same is true for the preservation of the black hole
mass outside the horizon. A finer analysis, however, shows that
the outer boundary conditions do not preserve the constraints and
this affects the consistency of the long term results. In spite of
that, the code runs forever (although we obviously stopped it when
all the numerical grid fell into the hole).

\section{Space plus time decomposition}
We will use the well known 3+1
decomposition\cite{Lich,Choquet,ADM} of the space-time metric,
namely
\begin{equation}\label{metric}
ds^2 = -\alpha^2\;dt^2 +
    \gamma_{ij}\;(dx^i+\beta^i\;dt)\;(dx^j+\beta^j\;dt) \;,
\end{equation}
where the lapse function $\alpha$ and the shift $\beta^i$ are
related to the choice of the space-time coordinates and
$\gamma_{ij}$ is the induced metric on every time slice. The
extrinsic curvature $K_{ij}$ (second fundamental form) of the
slices is given by
\begin{equation}\label{Kij}
    (\partial_t - {\cal L}_\beta)\, \gamma_{ij} = -2\alpha\,K_{ij}  \;,
\end{equation}
where ${\cal L}$ stands for the three-dimensional Lie derivative
operator.

Four of the ten Einstein field equations do not contain time
derivatives when expressed in terms of the previously defined
quantities. These four constraint equations can be easily
identified: the ``energy'', or Hamiltonian constraint
\begin{equation}\label{energy_constraint}
    2\alpha^2 \; G^{00} = {}^{(3)}R - tr(K^2) + (tr\,K)^2 \;,
\end{equation}
where $^{(3)}R$ is the trace of the three-dimensional Ricci
tensor, and the ``momentum constraint''
\begin{equation}\label{momentum_constraint}
    \alpha \; G^0_{\;i} = \nabla_k\;K^k_{\;i} - \partial_i(tr\,K) \;,
\end{equation}
where index contractions and covariant derivatives are with
respect to the induced metric $\gamma_{ij}$.

The time evolution of $K_{ij}$ is given by a system obtained from
the remaining Einstein field equations. For instance, the space
components of the four-dimensional Einstein tensor can be
written\cite{PRD97}
\begin{eqnarray}\label{Einstein}
    (\partial_t &-& {\cal L}_\beta)\, K_{ij} \;=\; -\nabla_i\alpha_j
    +\nonumber \\
    &&\alpha\; [{}^{(3)}R_{ij} -2K^2_{ij} + tr\,K\;K_{ij} - G_{ij}]\;
    \nonumber
    \\&& -\alpha/4\;\gamma_{ij}\;[{}^{(3)}R - tr(K^2) + (tr\,K)^2 -
    2\;tr\,G]\;,
\end{eqnarray}
where $^{(3)}R_{ij}$ stands for the three-dimensional Ricci
tensor. The matter terms in the perfect fluid case can be computed
from
\begin{equation}\label{matter}
    G_{ij} = 8\pi\,[(\mu+p) u_iu_j + p\,\gamma_{ij}] \;\; ,
\end{equation}
where $\mu$ is the total energy density of the fluid, $p$ is the
pressure, and $u_{i}$ is its fluid 3-velocity.

We call this set of equations (\ref{Kij},\ref{Einstein}) the
Einstein evolution system. Notice that it differs from the
standard 3+1 evolution system (obtained from the space components
of the four-dimensional Ricci tensor) by a term containing the
Hamiltonian constraint. This ensures that the matter contribution
(\ref{matter}) vanishes in the Newtonian limit. Even in the vacuum
case, the Einstein evolution system  has proved to be superior
when following the long term evolution for a spherically symmetric
(1D) black hole\cite{PRD97}.

\subsection{The momentum constraint as an evolution equation}
Let us consider now the quantities $V_i$, which can be obtained
from the metric first derivatives as follows
\begin{equation}\label{vector}
    V_i \equiv 1/2\;\gamma^{rs}(\partial_i\gamma_{rs}
                          -\partial_r\gamma_{is})\;.
\end{equation}
One can alternatively compute the time evolution of $V_i$ by using
the momentum constraint (\ref{momentum_constraint}) as a dynamical
equation, that is\cite{PRD97}:
\begin{eqnarray}\label{evolution_V}
    \partial_t V_i &-& \beta^k\partial_k V_i
    - (\partial_i \beta^r)\;V_r +\nonumber\\&&
    1/2\;(\gamma^{kr}\;\gamma_{is}-\delta^k_i\;\delta^r_s)\;
    \partial^2_{kr}\,\beta^s \;=\;\nonumber\\&&
    \alpha^2\;G^0_{\;i} + \alpha_r\;K^r_{\;i} - \alpha_i\;tr\,K
    - \nonumber\\&& \alpha\;[2\,K_i^{\;r}\;V_r + K^{rs}\;\Gamma_{irs}
               - K_i^{\;r}\;\partial_r(ln\sqrt{\gamma})].
\end{eqnarray}
We have chosen this second approach, which amounts to consider the
following set of basic dynamical quantities
\begin{equation}\label{full vars}
\{\gamma_{ij}\;, \;\;K_{ij}\;, \;\;V_i\} \;,
\end{equation}
so that the condition (\ref{vector}) can be now considered just as
an algebraic constraint which will hold if and only if the
momentum constraint is satisfied.

The three-dimensional Ricci tensor can be written in terms of
these quantities as
\begin{eqnarray}\label{Ricci}
    {}^{(3)}R_{ij} &=& -1/2\;\gamma^{rs}\,\partial^{\,2}_{rs}\,\gamma_{ij}
    + \partial^{\,2}_{ij}\;ln\sqrt{\gamma} - \partial_i\,V_j
    - \partial_j\,V_i + \nonumber\\&& \Gamma^k_{ij}\,(2\,V_k -
    \partial_k\,ln\sqrt{\gamma})
    +\gamma^{kl}\,\gamma^{rs}\,[(\partial_k\,\gamma_{ri})
    (\partial_l\,\gamma_{sj}) \nonumber\\&-& \Gamma_{ikr}\,
    \Gamma_{jls}] \;,
\end{eqnarray}
so that its trace can be easily expressed as follows
\begin{equation}\label{trR}
    {}^{(3)}R = -2\,\partial_k\,V^k + \Gamma_{krs}\;\Gamma^{rks}
    - \gamma^{rs}\;(\partial_r\,ln\sqrt{\gamma})
                 \;(\partial_s\,ln\sqrt{\gamma})\;.
\end{equation}

\subsection{Gauge conditions}
So far we have considered a completely general coordinate system.
We will keep our freedom to fix the shift components $\beta^i$ as
purely kinematical quantities. The choice of the lapse function
$\alpha$, on the contrary, will be closely related with the
dynamics of our system.

A popular choice is to impose the maximal slicing
condition\cite{York72}
\begin{equation}\label{maximal}
    tr\,K = 0\;,
\end{equation}
which amounts to compute $\alpha$ by solving the elliptic equation
obtained by taking the trace of (\ref{Einstein}) or a combination
of this trace with the Hamiltonian constraint
(\ref{energy_constraint}). This choice has proven to be quite
effective in some cases, although it is not free from long term
instabilities. Also, elliptic equations are very expensive in
terms of computational resources and getting high resolution long
term calculations on three-dimensional grids seems to be beyond
the capabilities of present day computers.

A different strategy is to link the evolution of the lapse
$\alpha$ to that of the volume element $\sqrt{\gamma}$ by means of
the following equation
\begin{equation}\label{lapse}
    (\partial_t - {\cal L}_\beta)\, ln\,\alpha =
    -\alpha\;f\;tr\,K ,
\end{equation}
where $f$ is a given non-negative function of $\alpha$. This opens
the way to many different choices of the time slicing. The
geodesic slicing is included as a subcase with $f=0$. The $f=1$
case corresponds to the ``harmonic
slicing''~\cite{Choquet83,PRD88} (the resulting time coordinate is
harmonic). Another interesting case is obtained when $f=1/\alpha$;
it mimics maximal slicing near a singularity, when the lapse
collapses to zero (a very similar case was considered
in~\cite{Bernstein,2DBH}. The related choices $f=n/\alpha$
($n=2,\,4$) mimic both the singularity-avoidance and the
large-gradients-avoidance properties of the maximal slicing for a
fraction of the computational cost.

So far we have obtained a closed dynamical system (once the shift
$\beta^i$ is given) which can be considered as the second-order
equivalent of the first-order hyperbolic system given in
Ref\cite{PRL95,PRD97}. The causal structure of this second-order
system will be discussed elsewhere\cite{Project}. All we can say
by now is that the direct equivalence with a first-order system
which is known to be hyperbolic seems a good starting point for
numerical relativity applications. But this feeling should be
confirmed by experience, as we will see in what follows.

\section{A robust evolution system}
\subsection{Keeping apart trace and trace-free components}
Both from the theoretical and from the practical point of
view\cite{Lich,York71}, it is clear that $tr\,K$ is directly
related to the gauge degrees of freedom, which should be treated
very carefully to avoid spurious 'gauge pathologies' or riddles
that do appear too frequently in 3D calculations\cite{MiguelJoan}.
This suggests the convenience of evolving separately the trace and
trace-free components of $K_{ij}$.

We will follow here the notation of Shibata and Nakamura\cite{SN}
to write down the following conformal decomposition of the space
metric
\begin{equation}\label{conf_metric}
    \gamma_{ij} = e^{4\Phi}\;\widetilde{\gamma}_{ij}
\end{equation}
with $\Phi$ chosen so that the determinant $\widetilde{\gamma}$ of
the conformal metric is equal to unity, namely
\begin{equation}\label{det}
    \sqrt{\gamma} = e^{6\Phi}\;.
\end{equation}
This means that we can split out the evolution equation
(\ref{Kij}) into the following pieces\cite{Lie_derivs}:
\begin{equation}\label{evol_gij}
    (\partial_t - {\cal L}_\beta)\, \widetilde{\gamma}_{ij}
    = -2\alpha\;A_{ij}  \;,
\end{equation}
\begin{equation}\label{evol_Phi}
    (\partial_t - {\cal L}_\beta)\, \Phi = -1/6\,\alpha\;K  \;,
\end{equation}
where we have noted for short
\begin{eqnarray}\label{K_pieces}
    A_{ij} &\equiv& e^{-4\Phi}\; (K_{ij} -1/3\,tr\,K\;\gamma_{ij})\;, \\
    K &\equiv& tr\,K\;.
\end{eqnarray}

We can also split out the evolution equations (\ref{Einstein},
\ref{evolution_V}) in the same way:
\begin{eqnarray}\label{evol_Aij}
            (\partial_t &-& {\cal L}_\beta) A_{ij} =
    e^{-4\Phi}\{ -\widetilde{\nabla}_i\alpha_j
    +2\;(\alpha_i\,\Phi_j+\alpha_j\,\Phi_i)\nonumber\\&
    +&1/3\;\widetilde{\gamma}^{rs}(\widetilde{\nabla}_r\alpha_s
    -4\;\alpha_r\,\Phi_s)\;\widetilde{\gamma}_{ij}
    \nonumber\\
    &+&\alpha \; [ {}^{(3)}R_{ij} - 1/3\;(\widetilde{tr}\,{}^{(3)}R)\;
     \widetilde{\gamma}_{ij} - \;G_{ij}\, ]\;  \}\nonumber\\&
   +& \alpha\;(K\;A_{ij}-2\, \widetilde{\gamma}^{rs}A_{ir}A_{js}
   )\;
\end{eqnarray}
\begin{eqnarray}\label{evol_K}
            (\partial_t &-& {\cal L}_\beta) K = e^{-4\Phi}\{
    -(\widetilde{\gamma}^{rs}\widetilde{\nabla}_r\alpha_s)
    - 2\;(\widetilde{\gamma}^{rs}\alpha_r\,\Phi_s)\nonumber\\&
    +& \alpha \; [1/4\;\widetilde{tr}\,{}^{(3)}R +
    1/2\;\widetilde{tr}\,G ]\;  \}
    \nonumber \\ &+& \alpha \;(1/2\;K^2 +
    3/4\;\widetilde{\gamma}^{rs}\;\widetilde{\gamma}^{kl}A_{kr}A_{ls})\;,
\end{eqnarray}
\begin{eqnarray}\label{evol_V}
            \partial_t V_i - \beta^k\partial_k V_i
    &-& (\partial_i \beta^r)\;V_r +
    1/2\;(\widetilde{\gamma}^{kr}\widetilde{\gamma}_{is}-\delta^k_i\;\delta^r_s)\;
    \partial^2_{kr}\,\beta^s \nonumber\\ = \alpha^2\;G^0_{\;i}
    &+&
    A_{ir}\widetilde{\gamma}^{rs}(\alpha_s-2\alpha\,V_s+2\alpha\,\Phi_s)
    \nonumber\\&-& \alpha\,\widetilde{\gamma}^{kr}\widetilde{\gamma}^{ls}
    A_{kl}\;\widetilde{\Gamma}_{irs}- 2/3\;K\;\alpha_i \;.
\end{eqnarray}

The resulting system (\ref{evol_gij}, \ref{evol_Phi},
\ref{evol_Aij}, \ref{evol_K}, \ref{evol_V}) allows to compute the
time evolution of the enlarged set of dynamical quantities
\begin{equation}\label{conf vars}
\{\widetilde{\gamma}_{ij}\;, \;\;\Phi\;, \;\;A_{ij}\;, \;\;K\;, \;
\;V_i\} \;,
\end{equation}
which has the same basic structure of the Einstein evolution
system (\ref{Kij}, \ref{Einstein}, \ref{evolution_V}) because the
two extra quantities we have introduced are redundant, so that the
constraints
\begin{equation}\label{conf_constraints}
    det \;\widetilde{\gamma}  = 1 \;, \;\; \widetilde{tr}\,A=0\;
\end{equation}
are trivially propagated in time.

\subsection{Numerical counter-terms}
In numerical applications the algebraic constraints
(\ref{conf_constraints}) can be easily monitored as error
indicators. Their behaviour is quite unstable: $\widetilde{tr}\,A$
grows without bound (the actual rate depends on every case) and
this makes $det \;\widetilde{\gamma}$ to go far away of its exact
value of $1$. This leads to an obvious inconsistency which causes
the numerical 3D code to crash after a relatively short time.

To fight against these errors, we have modified the
right-hand-sides of the evolution equations (\ref{evol_gij},
\ref{evol_Aij}) by adding the following counter-terms:
\begin{equation}\label{counter_det}
    (\partial_t - {\cal L}_\beta)\, \widetilde{\gamma}_{ij} =
    \;\cdots\; - 1/3\; (det \;\widetilde{\gamma} -1)/\tau
    \;\widetilde{\gamma}_{ij}\;,
\end{equation}
\begin{equation}\label{counter_tr}
    (\partial_t - {\cal L}_\beta)\, A_{ij} =
    \;\cdots\; - 1/3\; (\widetilde{tr}\,A)/\tau
    \;\widetilde{\gamma}_{ij}\;,
\end{equation}
where $\tau$ is a 'time constant'.

It is easy to see that the counter-terms (\ref{counter_det},
\ref{counter_tr}) by themselves would cause an exponential
decrease of the errors in $\widetilde{tr}\,A$ and $det
\;\widetilde{\gamma}$. In practice, we have set the time constant
$\tau$ to the same value of the numerical time step $dt$ in order
to keep these errors within an small order of magnitude.

One should notice at this point that these counter-terms do vanish
in the continuum limit (at least for second-order-accurate
numerical algorithms), so that they do not affect at all the
structure of the analytical evolution system. Harmless from the
theoretical point of view, they provide a valuable tool that
allows us to perform robust 3D calculations, as we will see below.
See Refs. \cite{Waimo,Eppley} for the use of numerical control
terms in elliptic gauge conditions.

\section{Application to the Black Hole case}
 Let us start with the usual time symmetric
initial data ($K_{ij}=0$) for a conformally flat metric (space
isotropic coordinates), corresponding to a Schwarzschild black
hole. Let us now replace the interior region by a constant density
incoherent matter content, keeping the exterior vacuum region of
the original hole\cite{stuffed}. This 'stuffed' black hole
approach allows one to start from singularity-free initial data
and avoids then the usual procedure of excising the inner region
from the computational grid, which would introduce an internal
boundary at the black hole horizon.

\begin{figure}
\centerline{
\psfig{figure=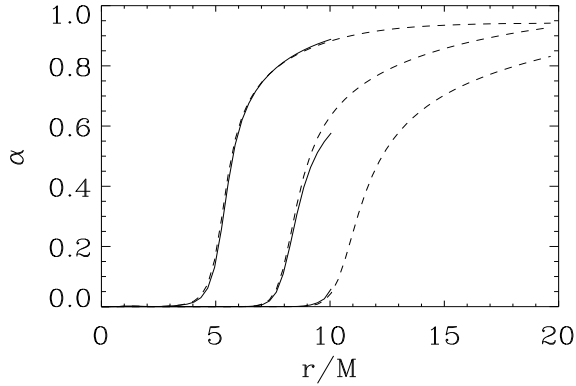,height=60mm,width=90mm}}
\vspace{0.2in} \caption{The evolution of $\alpha$ is compared for
two positions of the external boundary ($r=10M$ and $r=20M$) for a
Schwarzschild black hole. The resolution is set at $dx=0.3M$. The
lapse is plotted at $t=27M$, $37M$ and $47M$. In the latter the
black hole has gone almost out of the smallest grid. We can see
that the agreement is quite good.} \label{alpbound}
\end{figure}

We have performed our calculations in the Cactus code\cite{Cactus}
environment. In order to save computational resources, we have
taken advantage of the equatorial symmetry of the problem to
evolve just an octant of the space slices. This requires to
introduce boundary conditions at the symmetry planes. The outer
boundary has been treated as in Ref\cite{SN}. In
Fig.~\ref{alpbound} we consider the effect of this boundary
condition over the collapse of the lapse. The results seem to be
quite independent on the position of the outer boundary.

\begin{figure}
\centerline{
\psfig{figure=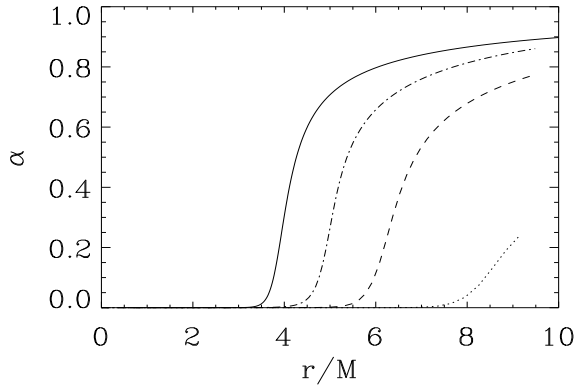,height=60mm,width=90mm}}
\vspace{0.2in} \caption{The convergence of the lapse towards 1D
results is shown. $\alpha$ is plotted against $r/M$ at $t=30M$ for
a Schwarzschild black hole. The solid line corresponds to the 1D
result with high resolution. The remaining lines, converging to
that solution, correspond to 3D results with $dx=0.3M$, $0.15M$
and $0.075M$.} \label{alpconv}
\end{figure}

As coordinate conditions, we use normal coordinates ($\beta^i=0$)
and an algebraic slicing of the form (\ref{lapse}) with the choice
$f=2/\alpha$ which leads to smoother profiles of the evolved
quantities. To start a dynamical simulation we can take as usual a
constant initial lapse. The time evolution is then obtained by
using the 'icn' second-order-accurate algorithm which has been
incorporated to the Cactus version currently used by the Potsdam
group. The actual plight for the reliability of this collapse
turns out to be resolution, as we can see in Fig.~\ref{alpconv}.
The 3D results converge to the well established 1D solution and we
see that a resolution finer than $0.1M$ is needed if one wants to
get really close to the physical solution.

\begin{figure}
\centerline{
\psfig{figure=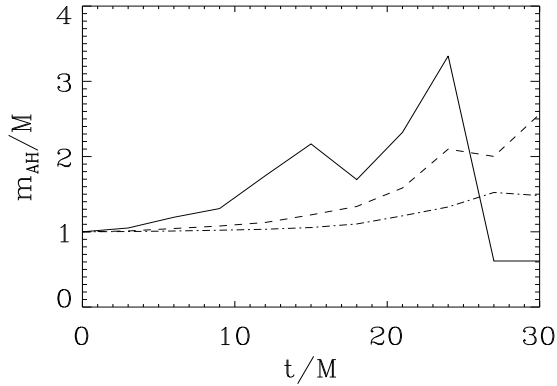,height=60mm,width=90mm}}
\vspace{0.2in} \caption{The Bondi mass at the apparent horizon is
plotted against time for a Schwarzschild black hole. Results with
three resolutions ($dx=0.3M$, $0.15M$ and $0.075M$) are plotted,
which show convergence to the analytical value $m_{AH}/M=1$.}
\label{massconv}
\end{figure}

As far as our test-bed problem has spherical symmetry, we can make
use of the Bondi mass function to monitor the convergence of the
numerical solution towards the analytical one. In
Fig.~\ref{massconv} we plot the time evolution of the numerically
obtained mass at the apparent horizon. The results show the
expected convergence to the constant line corresponding to the
analytical value. This confirms our previous conclusions.

\begin{figure}
\centerline{
\psfig{figure=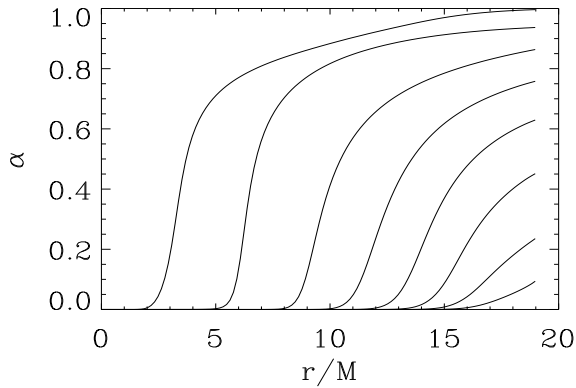,height=60mm,width=90mm}}
\vspace{0.2in} \caption{The collapse of the lapse for a
Schwarzschild black hole is shown. $\alpha$ is plotted against the
radius for increasing times ($t=15M$, $30M$, $45M$, $60M$, $75M$,
$90M$, $105M$, $120M$). The whole grid goes eventually inside the
hole and the calculation is stopped (although it never crashes).
At late times, however, the numerical solution departs from the
physical one due to boundary effects.} \label{alplong}
\end{figure}

Finally, we show in Fig.~\ref{alplong} the performance of the code
in long runs. We can follow the black hole evolution for unlimited
time. We just stop the calculation when the numerical grid is
almost completely inside the hole.

\section{Conclusions}
The equations we propose are interesting both from the theoretical
and for the numerical point of view. The decomposition of $K_{ij}$
into trace and trace-free parts allows a separate monitoring of
the numerical errors which leads to the following quite surprising
conclusion: the numerical trace of the trace-free part becomes
easily unstable in strong field 3D calculations. Notice that this
decomposition is not made in most Numerical Relativity formalisms,
so that the errors arising from the trace-free part are directly
mixed up with the dynamics of the trace part. This might explain
the arising of the so called 'gauge
pathologies'\cite{Miguel,MiguelJoan} even in situations with
apparently harmless initial data.

The counter-terms we have introduced allow one to deal with this
problem and this is enough to perform robust 3D Numerical
Relativity calculations, even in the black hole case. We have
demonstrated this by actually running the code up to times greater
that one hundred black hole masses without any sign of
instability: we just stopped when nearly all the numerical grid
was inside the hole.

\acknowledgments
 This work is supported by the Direcci\'on General para
Investigaci\'on Cient\'{\i}fica y T\'ecnica of Spain under project
PB97-0134 and the Conselleria d'Educaci\'{o}, Cultura i Esports of
the Balearic Government under an A7 grant.

We want to thank the Numerical Relativity groups of the Albert
Einstein Institute in Potsdam and the Washington University
(headed by Edward Seidel and Wai-Mo Suen, respectively) for useful
discussions and for allowing comparison with their own data.
Special thanks are due to Miguel Alcubierre and Bernd Bruegmann
for providing access to limited access software tools in the
Cactus code environment which have been used to speed up our code
development.

\bibliographystyle{prsty}

\begin{thebibliography}{99}
\bibitem{Lie_derivs} Notice that
    $\Phi$,$\widetilde{\gamma}_{ij}$,$A_{ij}$ are three-dimensional
    tensorial densities (not tensors!) and the corresponding Lie
    derivatives must be computed accordingly.
\end{thebibliography}

\end{document}